\documentclass[useAMS,usenatbib]{mn2e}
\usepackage[dvips]{graphicx}
\usepackage{amssymb}
\usepackage{txfonts}
\usepackage{lscape}

\newcommand{\thissource}{SAX J1712.6-3739}

\newcommand{\ha}{H\,${\alpha}$}

\newcommand{\ch}{{\it Chandra}}
\newcommand{\xmm}{{\it XMM-Newton}}
\title[A bow shock nebula associated with LMXB SAX J1712]{Discovery of a large and bright bow shock nebula associated with low mass X-ray binary \thissource}
\author[K. Wiersema et al.]{K. Wiersema$^{1}$\thanks{E-mail:
kw113@star.le.ac.uk},   D.~M.~Russell$^{2}$, N.~Degenaar$^{2}$, M.~Klein-Wolt$^{2,3}$, R.~Wijnands$^{2}$, \and S.~Heinz$^{4}$, A.~M.~Read$^{1}$, 
 R.~D.~Saxton$^{5}$, N.~R.~Tanvir$^{1} $\\
$^{1}$ Department of Physics and Astronomy, University of Leicester,  Leicester LE1 7RH, UK\\
$^{2}$ Astronomical Institute ``Anton Pannekoek'', University of Amsterdam, Kruislaan 403, 1098 SJ Amsterdam, The Netherlands \\
$^{3}$ ALTRAN, De Fruittuinen 30, 2132 NZ Hoofddorp, The Netherlands\\
$^{4}$ Department of Astronomy, University of Wisconsin-Madison, 6508 Sterling Hall, 475 North charter Street, Madison, WI 53593, USA\\ 
$^{5}$ ESA/ESAC, Apartado 78, 28691 Villaneuva de la Ca\~{n}ada, Madrid, Espa\~{n}a
}
\begin{document}

\date{Accepted 2009 February 17. Submitted 2008 December 22.}

\pagerange{\pageref{firstpage}--\pageref{lastpage}} \pubyear{2008}

\maketitle

\label{firstpage}

\begin{abstract}
In a multiwavelength program dedicated to identifying optical counterparts of faint persistent X-ray sources in the Galactic Bulge, we
find an accurate X-ray position of \thissource\ through \ch\ observations, and discover its faint optical counterpart using our data from EFOSC2 on the ESO 3.6m telescope.  We find this source to be a highly extincted neutron star LMXB with blue optical colours. We serendipitously discover a relatively bright and large bow shock shaped 
nebula in our deep narrowband \ha\ imaging, most likely associated with the 
X-ray binary. A nebula like this has never been observed before in association with a LMXB, 
and as such provides a unique laboratory to study the energetics of accretion and jets. We put forward different models to
explain the possible ways the LMXB may form this nebulosity, and outline how they can be confirmed observationally.
\end{abstract}


\section{Introduction} \label{sec:intro}
The bright - X-ray luminosities in the order of 10$^{36}$ erg s$^{-1}$ and higher - accreting 
black hole and neutron star X-ray binary systems are amongst the best studied Galactic X-ray sources.
Many of these show a rich phenomenology, displaying spectral state changes, dramatic accretion outbursts and
complex temporal behaviour on a wide range of temporal and spectral scales.  
However, an interesting sub-class of X-ray sources exists that are {\em persistent} in their X-ray behaviour, but have low X-ray luminosities. 
The monitoring program of the Galactic bulge source population with the {\em RXTE} satellite through PCA scans 
started in 1999 and has revealed many new bright X-ray transients. It also detected a large number of these faint, persistent sources, for which relatively little is known.

We have undertaken a small survey using \ch\ and the ESO 3.6m telescope to accurately localize a sample of 20 of these {\em RXTE} faint persistent X-ray source
sources, and determine their nature: this class of sources is thought to harbour a wide variety of
exotic types of systems (e.g. the class of the very short binary period sources - ultracompact X-ray binaries). The main results of this survey will be presented in a forthcoming paper (Wiersema et al in prep; hereafter W09). 

One of the sources of this survey was \thissource, a candidate ultra-compact X-ray binary (UCXB;  
\citealp{intzand2}).  In the case of a UCXB  the mass accretion rate 
or luminosity may be a good indicator for the mass transfer rate from the donor star, and such low accretion rates 
suggest the possibility of degenerate donors such a white dwarfs, making UCXBs important sources to study from
a binary evolution point of view. However, their (optical) faintness makes UCXBs particularly hard to find and unambiguously identify (see for example \citealp{nelemans}).

At optical wavelenghts, several components contribute to the light received from a LMXB, e.g. the accretion disk, the secondary star
 and a possible jet component. To broadly separate these various contributions and eliminate other,
perhaps less likely, possible source identifications (e.g. accreting white dwarfs, high mass X-ray binaries; see W09 for statistics from our survey), we observed 
the sources in the survey of W09 in four optical bands. 

In our narrowband  \ha\ imaging of \thissource, we discover a large, bright, roughly bow shock-shaped nebulosity, with an origin consistent with the
position of the \ch\ counterpart of \thissource, and unlike any other previously discovered nebulosity associated with an X-ray binary.
The importance of this discovery is clear: the production of the \ha\ nebula in \thissource\ is undoubtedly related to the accretion produced photon flux or the 
mechanical work of an accretion-associated jet slamming into the ISM (and the kinetic energy of the binary), and in both cases the morphology and flux of the nebulosity can be used to constrain the current and past energy output of this candidate neutron star UCXB. 
%The fact that this source is a candidate ultracompact neutron star X-ray binary makes this an even more exciting possibility.

\section{X-ray observations}\label{sec:xray_obs}
As part of the survey described in W09 we observed the {\em BeppoSAX} position of \thissource\   
with the High Resolution Camera (HRC-I) onboard \ch, on  2007 October 23, with an effective on-source exposure time of 1.176 ks. 
The data were processed using the CIAO tools (version 4.0) and standard \ch\ analysis threads.
Using the {\em wavdetect} routine, a bright point source is found with count rate 5.16 counts s$^{-1}$, at position RA: 258.15319 degrees, Dec:  -37.64473 degrees (J2000), 
with an astrometric uncertainty of 0.6 arcseconds (90\% containment, including systematics). This position is within the errorcircle of the {\em ROSAT} position of \thissource, the
best previous position of this source. Based on position and observed count rate we conclude that the \ch\ source can be identified with \thissource.

\begin{figure}
\centerline{\includegraphics[width=8cm]{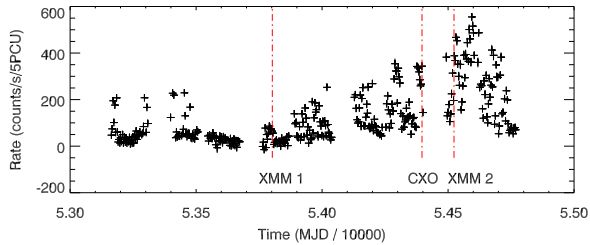}}
\caption{The PCA bulge scan lightcurve spanning several years, taken from the PCA Bulge scan web database. Indicated by dashed lines are the times of our \ch\ observation and the two \xmm\ slews. } 
\label{fig:pcalc}
\end{figure}

We find that the \xmm\ slew survey \citep{xmmslew} lists a bright source consistent with the \ch\ position: XMMSL1 J171236.3-373841. The source is detected in two slews: the first during  2006 March 8 and the second during 2008 February 27, at position RA: 258.15194 degrees,  Dec: -37.644827 degrees (J2000; $\sim6"$  error).
\xmm\  EPIC-pn 0.2--12 keV  count rates were high:  23 counts s$^{-1}$ in slew 1, and  $64$ counts s$^{-1}$ in slew 2, with effective exposure times of 3.7 
and 4.2 seconds, respectively. 
PCA observations at similar times also show a  count rate difference of a factor $\sim3.5$ (Figure \ref{fig:pcalc}), strengthening the identification of the \xmm\ source with the PCA source even more.

 \cite{cocchi} use the {\em BeppoSAX} data to find $N_{\rm H}  \sim1 - 2  \times 10^{22}\,\, {\rm cm}^{-2}$, and $\Gamma \sim 2.2$. 
We extract spectra from the \xmm\ slew data, and fit them with an absorbed powerlaw, where we fix $\Gamma = 2.2$ in order to get a good estimate of the
absorption, needed to interpret the optical data (see Section \ref{sec:lmxb}), as there are not sufficient counts to fit for both parameters. 
We find $N_{\rm H} = 1.89^{+1.11}_{-0.66} \times 10^{22}\,\, {\rm cm}^{-2}$ (90\%)
for slew 1, and $N_{\rm H} = 2.36^{+0.56}_{-0.47} \times 10^{22}\,\, {\rm cm}^{-2}$ (90\%) for slew 2. This is consistent 
with, though on the high end of,  the findings of \cite{cocchi}.

\begin{figure}
\centerline{\includegraphics[width=8cm]{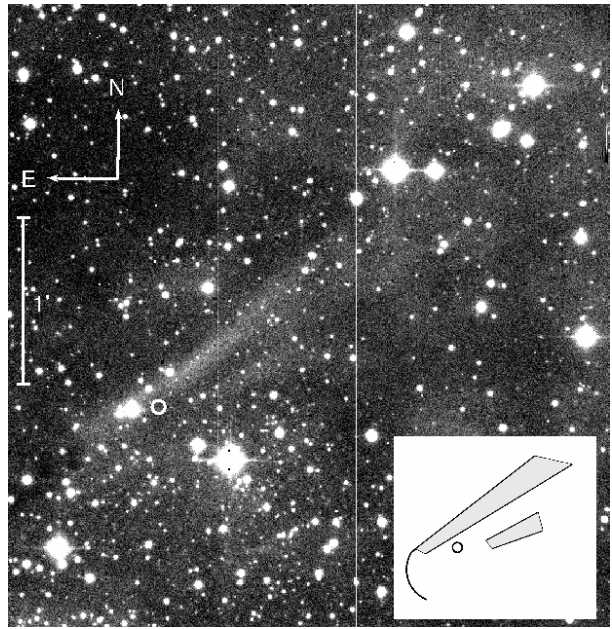}}
\caption{EFOSC2 \ha\ image of the field of SAX J1712.6-3739, excluding the areas on the sides that are most affected by vignetting. The position of the bright \ch\  source is indicated with a white circle. Note that this circle is much larger than the
\ch\ errorcircle, for presentation purposes.  The inset gives a rough guide to the main features. } 
\label{fig:nebulaSAX}
\end{figure}

\section{Optical observations}\label{sec:optical_obs}
Observations were performed in Service Mode with the EFOSC2 instrument on the ESO 3.6m telescope at La Silla, Chile,
at the {\it BeppoSAX} position of \thissource, as part of the survey described in W09.
This configuration gives a field of view of $\sim4.5 \times 4.5$', with a pixel scale of  
 0.31" per pixel on sky ($2 \times 2$ binning).
Data were taken on 2007 July 1, in the following filters: Bessel V (ESO filter \#641; total exposure time 10 minutes),
Bessel R (\#642; 10 minutes), Gunn i (\#705; 10 minutes) and narrowband \ha\ (\#692, 15 minutes). 
Conditions cannot be considered photometric with certainty, and as such an accurate absolute calibration using photometric standard stars cannot be performed.
A full log of the optical observations can be found in Table \ref{table:opticallog}.

All data were overscan corrected, trimmed, flatfielded and cosmic ray cleaned with the usual procedures, using tasks in IRAF. 
Data were taken close to the Moon, giving rather high sky background, resulting in somewhat lower limiting magnitudes for these bands than can be expected under the 
seeing and airmass conditions given in Table \ref{table:opticallog}. 

WCS calibration of the optical frames is carried out by matching sources in the frames to USNO-B2 source lists. The resulting WCS solutions (RMS 0.3$"$)  were subsequently refined using the Two Micron All Sky Survey (2MASS) sources detected in the field.
Resulting RMS values of the best astrometric fit for each filter are 0.13$"$ for the R and i band frames (using 590 2MASS stars), and 0.16$"$ for the V and \ha\ frames (400 stars).

As this field is in the Galactic plane (latitude $b = 0.92$ degrees) the field is crowded, particularly in i and R band. 
 We therefore used IRAF {\sc DAOPHOT} tasks to perform iterative point spread function (PSF) fitting photometry: after detecting a bright sample of stars with DAOFIND,
we fit a first order spatially variable empirical PSF to bright stars (using several iterations to subtract off neighbour stars, refine PSF star selection
and use progressively larger fit radii). 
After PSF photometry using ALLSTAR we subtract this list of stars and iterate with a lower detection threshold.
 
 EFOSC2 is a focal reducer instrument, and as such suffers from a small amount of vignetting, and the
PSF for sources near the chip edges is distorted. The position of the optical counterpart of \thissource\   is sufficiently away from the 
chip edges that this does not affect our results (see Figure \ref{fig:nebulaSAX}).

To derive approximate magnitudes, we use EFOSC2 zeropoints and La Silla average atmospheric extinction terms from the EFOSC2 Quality Control webpages, 
calibrating the Gunn i band data to I. All magnitudes are given in Vega magnitudes. 
While the absolute photometry is likely not highly accurate, the differential colours of objects in the field should be reasonably reliable, allowing a 
classification of sources. Note that due to the low Galactic latitude no SDSS photometry is available for field star calibration.

We use the V, R, I and \ha\ photometry lists to construct colour-magnitude diagrams (CMD) and colour-colour diagrams (CCD).
For each colour magnitude diagram point sources are matched for the relevant filters only. The R -- \ha\ colour cannot be reliably quantified, as no spectrophotometric standard was observed. However, most sources form a narrow strip of 
 R -- \ha\ colour (i.e. no unusally strong \ha\ emission or absorption with respect to the R band continuum), which we shift to an average value of 
 R -- H\,$\alpha \sim 0$, to guide the eye. As the \ha\ line is not situated directly in the center of the Bessel R band, we use a R -- i vs R -- \ha\ diagram whenever possible,
 following the procedure outlined in Witham et al. (2006) to identify \ha\ emitters near the \ch\ position.
To find potential optical counterparts to the X-ray source, we add the (90\%) RMS of the WCS calibration of the optical images to 2MASS in quadrature to 
the (90\% containment) X-ray position error for each filter, and search for optical point sources within the resulting error circle (Figure 3).

\begin{table}
 \centering
 \scriptsize
 %\begin{minipage}{140mm}
  \caption{Log of our observations of \thissource. The 5$\sigma$ limiting magnitude listed is derived for the 
  {\em entire} EFOSC2 frame, and is therefore not equivalent to a detection limit of a specific source on a specific place on the chip (i.e. when confusion / blending plays a role). Seeing is
  measured from the PSF on the frames, excluding the areas affected by vignetting. }\label{table:opticallog}
  \begin{tabular}{@{}lllllll@{}}
  \hline
                 & Band & Observing time         &  Exp. time & Seeing  & Airmass & Lim. mag.  \\ 
                                       &            & start  (UT)                  &  (s)                      & (arcsec) &                & (5 $\sigma$)                           \\
 \hline
  EFOSC2                    &      R   & 01/07/2007 04:28  & $2\times 300$   &     0.65         &  1.04   &        23.4                       \\         
                                       &     V   & 01/07/2007 04:39   & $2\times 300$   &     0.76         &  1.06   &         23.5                          \\                                  
                                       &      I    & 01/07/2007 05:51  & $2\times 300$   &      0.85         &  1.07   &         23.0                      \\
                                       &   \ha\ & 01/07/2007 05:02  & 900                      &      0.71         &  1.08   &          -                 \\
 \hline
 \ch\                  & HRC-I            &     23/10/2007                               &  1.176 ks                    &               &              &       \\
 \hline
\end{tabular}
\normalsize
%\end{minipage}
\end{table}

\begin{table}
 \centering
 \scriptsize
 %\begin{minipage}{140mm}
  \caption{Results on the optical counterparts of \thissource. The optical positions are measured on the I band frame, where S/N is greatest, and include the
  absolute WCS calibration uncertainty to the 2MASS astrometry (see Section \ref{sec:optical_obs}).  Please note that the night was likely not photometric, and therefore
  the errors given on the magnitudes below do not include an error from absolute photometric calibration. }\label{table:results}
  \begin{tabular}{@{}llllllll@{}}
  \hline
          & Position (J2000) & Error ($"$)& I & R         &  V    \\ 
 \hline
  S1               &  258.15318 -37.64501 &0.25& 21.45 $\pm$ 0.06 & 22.61 $\pm$ 0.12 &    23.93 $\pm$  0.26               \\        
  S2                & 258.15376 -37.64465 &0.25 & 21.38 $\pm$ 0.04 & 23.05 $\pm$ 0.13 &    $>24.1$  (3$\sigma$)                  \\                                  
 \hline
\end{tabular}
\normalsize
%\end{minipage}
\end{table}

\section{SAX J1712.6-3739} \label{sec:lmxb}
This source has been discovered in 1999 (in 't Zand et al 1999), and is a persistent source, detected by {\em ROSAT}, {\em BeppoSAX} and {\it INTEGRAL}
(alternative identifiers {\it INTEGRAL}1 47 and 1RXS J171237.1-373834). It has been classified as
a neutron star LMXB, on the basis of several type I X-ray bursts, which set the distance at 6 - 8 kpc  (\citealp{cocchi}; see e.g. in 't Zand et al. 2007 for a discussion on past observations).

 \begin{figure}
\centerline{\includegraphics[width=8cm]{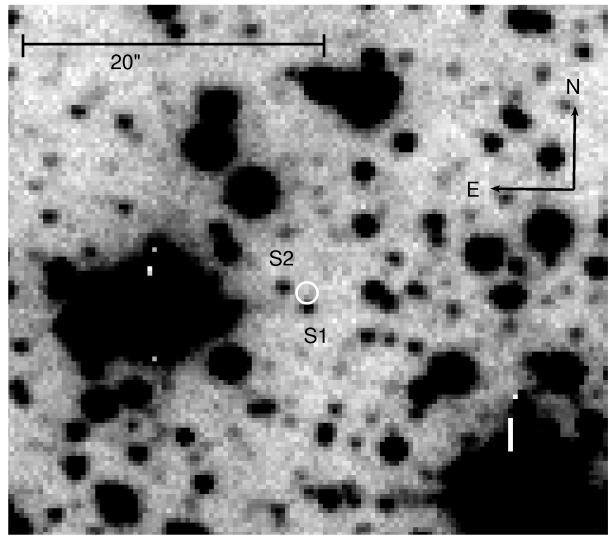}}
\caption{A small section of the EFOSC2 i band image of the field of SAX J1712.6-3739. The position of the bright \ch\  X-ray source is indicated with a white circle.  
Two optical sources are located close to the \ch\ position, labelled S1 and S2. S1 is located significantly closer to the \ch\ position and has bluer colours than S2.
 } 
\label{fig:finderSAX1712}
\end{figure}

 The EFOSC2 \ha\  data shows patchy nebulosity, which is confirmed through much larger field of view \ha\ data retrieved from the SuperCosmos repository (\citealp{supercosmos}): 
 the X-ray source is located in an area abundant in bright diffuse clouds of gas. 
 In the EFOSC2 data  a bright nebular source is obvious, which appears distinctly jet or bow shock shaped and spans most of the \ha\ image (see Figure \ref{fig:nebulaSAX}). 
The R band data are, not surprisingly, the only wideband data showing the central part of this nebula, consistent with an (\ha) line emission dominated nebula.

\begin{figure}
\centerline{\includegraphics[width=8cm]{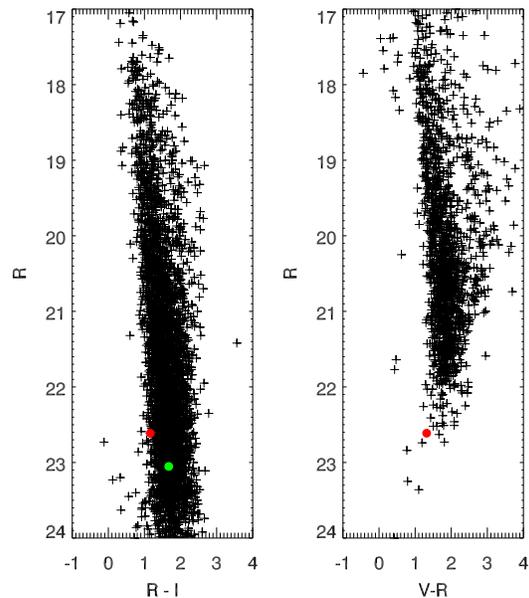}}
\caption{CMDs of the field of \thissource, with the optical candidates S1 indicated in red, and S2 in green. Errorbars are omitted for clarity. S1 is clearly a blue source, and is one of the faintest R band sources with a V band detection.
S2 is not detected in V. }%Neither source is detected in the \ha\ data, but limits are not constraining.} 
\label{fig:cmdSAX}
\end{figure}

The optical data shows two point sources very close to the \ch\ position, both detected in R and i, see Figure \ref{fig:finderSAX1712} and Table \ref{table:results}. The more southern source (hereafter 
called S1) is marginally detected in V ($\sim4 \sigma$), somewhat hindered  by 
the proximity to diffraction spikes of a brighter star. The other source (hereafter called S2), to the East of the \ch\ position is not detected in V.
The position of S1 (see Table \ref{table:results}) is close to the X-ray position (1.0$"$ from the center of the \ch\ errorcircle, i.e. consistent within $ 2\sigma$), whereas S2 is considerably further away (2.1$"$).

At the positions of these two sources no point source is found in the
\ha\ frame.  Using the method by \cite{witham} to plot sources in a R - \ha\ vs R - I diagram to select \ha\ emitters, we find that 
the \ha\ detection limits are not constraining for  either source. 
S1 has clearly very blue (intrinsic) colours,  see Figure  \ref{fig:cmdSAX}, consistent with expected colours of a LMXB (though note that this depends on the adopted extinction value, which is high and very uncertain).
Position and colours make S1 a more likely counterpart - though spectroscopy or photometric variability will be required to confirm the association.

\citet{intzand2} suggest from a literature data compilation that this source is a candidate UCXB. They derive a reddening by using the \citet{predehl} relation between Galactic $N_{\rm H}$ and $A_{\rm V}$, using $N_{\rm H} = 1.3 \times 10^{22}$ cm$^{-2}$ (\citealp{cocchi} and \citealp{intzand2}, a value roughly consistent with our estimate, see Section \ref{sec:xray_obs}) and find $A_{\rm V} = 7.3$. In combination with the distance of $\sim7$ kpc (Section \ref{sec:lmxb}), \citet{intzand2} predict a visual magnitude of $\sim26$ for a UCXB and $\sim22$ if the source is a non-ultracompact binary.
Sources S1 and S2 have V magnitudes (or limit in the case of S2) in between these two values, see Table \ref{table:results}, but obviously this is rather dependent on the uncertain extinction towards this source.  
Using this same extinction estimate and distance we find $M_{\rm V} \sim +2.4$ for S1. While this is faint, it is somewhat brighter than the absolute magnitudes of UCXB candidates, and comparable to standard short period LMXB values (van Paradijs \& McClintock 1994), though we note the uncertain optical extinction. 
Time resolved optical spectroscopy of this system would be able to reveal the binary period and therefore the UCXB nature.

 \section{The nature of the nebula}\label{sec:nebulanature}
 The \ch\ position coincides with the position from which the jet- or bow shock shaped nebula visible in \ha\  appears to originate, see Figure \ref{fig:nebulaSAX}, strengthening a connection between this object and the nebulosity. Specifically, an extension of the two approximately straight band /
stripes of \ha\ emission meet near the location of the LMXB. The probability of chance coincidence of a LMXB with an, unassociated,  bow shock shaped nebula can however not be robustly calculated, as it would
 require searching \ha\ surveys for similar features {\em unrelated} to LMXBs, while the definition of ``similar'' is necessarily influenced by the detection of this current, unique, source. 
It is important to emphasize here that a theoretical model predicting a system similar to this observed one (\citealp{heinz}) has been made {\em prior} to our discovery of  this system.

We may expect the LMXB to move roughly with the Galactic plane, as a large kick is required to make it move away from the plane.  
The projected direction of motion of the LMXB inferred from the bow shock shape is towards the Galactic plane, i.e. moving towards lower Galactic latitude, consistent with this picture.

In the following we continue on the premise that the nebula is indeed produced by the LMXB, and discuss the two most likely mechanisms to produce this nebula, discussing also which future observations may help distinguish these models. 

\subsection{Morphology}
The nebula seems to exist of two nearly parallel trails of \ha\ emission that appear to originate close to the LMXB, and a diffuse, near-semicircular region of emission just  `in front' (to the south-east) of \thissource. The distance from the LMXB to the leading edge of this diffuse region is $\sim30"$, which corresponds to a
 distance of $\sim3 \times 10^{18}$ cm (adopting $d = 7$ kpc). The trails appear to fade and widen with distance from the LMXB except initially; the maximum surface brightness of the trails is
 located at a distance $\sim20"$ and $\sim40"$ from the LMXB for the northern and southern trails, respectively. The width of each trail is $\sim9"$ close to the LMXB, widening to $\sim33"$ at a distance of $100"$, for the northern trail.
 %, which has position angle $\sim$307 degrees (from North to East); the lower one $\sim298$ degrees. 

\subsection{A jet-powered nebula?}
Accretion is on almost all masses and timescales observed to relate to the formation of high velocity jets, from the extremely energetic gamma-ray bursts to the more sedate
outflows of  young stellar objects. Evidence for jets in X-ray binaries is found from direct detection of fast moving radio or X-ray emission from the jets themselves or
shocks downstream (e.g. \citealp{tinget95,fomaet01,corbet02,tudoet08}), from transient (unresolved) radio emission (\citealp{fend01}), from spectral/polarimetric/timing
studies in the infrared and optical (e.g. \citealp{corbfe02,malzet04,russfj07,shahet08,russfe08}) or in one case from Doppler-shifted emission lines from baryonic jets
(\citealp{marget79}). Recently, bow shock nebulae powered by the jets of Cyg X--1 and LMC X--1 have been identified, and their properties have been used to infer the
time-averaged power of the jets (\citealp{gallo,russet07,cooket07}). 
Since \thissource\ has a hard spectrum and a steady luminosity we may assume it to have compact jets, detectable at radio frequencies, with typical flux densities $\sim0.1-1$ mJy (\citealp{mig}), unfortunately fainter than the limit of the NVSS survey locally. % ($\sim15$ mJy). 

Judging by the density of \ha\ clouds in the field (from the SuperCosmos data), this source is situated in a dense interstellar medium. The long term interaction of a jet from a LMXB with a dense medium coupled with a high
space velocity may give rise to bow shock-shaped trails of collisionally excited gas (see for details \citealp{heinz}). The requirement of a high space velocity means that
LMXBs should more easily form jet trails than HMXBs. However, sources with a morphology like this have not been observed to date, probably because for most LMXBs the local
ISM density is not high enough (\citealp{hein02}). The Heinz et al. (2008) model predicts a %nebula 
morphology and size remarkably similar to the observed \ha\ nebula (Figure
\ref{fig:nebulaSAX}) which would make \thissource\ the  first confirmed jet trail source. In this scenario, we interpret the nebula as an edge-brightened cone-shaped bow shock
(a `head-tail source', like those seen associated with AGN) powered by two continuously replenished jets. The space velocity of the LMXB would exceed the velocity of the
bow shock, forming trails behind the LMXB as opposed to shell-like stationary cocoons like the one of Cyg X--1 (\citealp{gallo}). The emission region `in front' of the LMXB is
expected in this scenario, ionized by the approaching radiation field. The \ha\ emission originates in this case from recombination of the shock-compressed gas near the shock
front. Radio plasma consisting of relativistic particles and magnetic fields should fill the region behind the bow shock front -- these trails may be detectable, as might
radio-emitting bremsstrahlung radiation from the shell. The detection of such structures will confirm the jet-powered nature of the nebula. 
When interpreting the nebula as a jet blown trail, the width of the trail provides important constraints on the system. In particular, using equation 12 from \cite{heinz} we find that the average jet power $W$ is likely well below 10$^{37}$ erg s$^{-1}$, not unexpected for a neutron star LMXB (though note that prototypical neutron star jet source Cir X-1 has a jet power close to 10$^{37}$ erg s$^{-1}$; see e.g. \citealp{tudoet08} and references therein).  In particular, the persistently active, neutron star UCXB 4U 0614+09 has been shown to produce powerful jets (\citealp{mig2}) with jet power $\sim10^{35}$ erg s$^{-1}$, dominating the 
radio to nIR emission from the system (\citealp{mig2, vrtilek}, Migliari et al in prep). The jet powered scenario is therefore not unexpected.

  \ha\ bow shocks with broadly similar morphology are occasionally found around fast moving isolated neutron stars, though much fainter and smaller than this source (e.g. \citealp{chatterjee}), powered by pulsar winds. 
Accretion inhibits strong pulsar winds in LMXBs but instead, accretion disc winds powered by irradiation on the disc            
could be powerful. However, disc winds are likely to be orders of magnitude weaker than the jets (see the case of 4U 0614+09 above), making it much more likely that the trails are powered by jets than by a disc wind.

\subsection{An X-ray-ionized nebula?}
Since \thissource\ is a persistent X-ray source, local nebulosity will be photo-ionized by its X-ray and UV radiation field. The nebula surrounding LMC X--1 is partly ionized
 in this fashion (\citealp{pakuan86,cooket07}). The morphology of the \thissource\ nebula may be consistent with this interpretation if the LMXB has a high
space velocity. 
However, the observed edge brightened morphology is hard to achieve in this scenario, while expected in the jet trail model. Furthermore we would need either a definite density gradient along the binary trajectory or a declining ionizing photon flux as a function of time to create the observed nebula morphology.
 
Emission line ratios are commonly used as diagnostics of shock-ionized or photo-ionized gas (e.g. \citealp{coxra85,alleet08}); for
example the [S\,II] / \ha\ line ratio is an indicator of shocked gas and the He\,II / \ha\ ratio of photo-ionized gas. Follow-up spectroscopy (or narrow-band imaging) of the trails will therefore constrain the nature of the nebula and probe the velocity, density and composition of
the gas, constraining either the time-averaged power of the jets or the soft X-ray/EUV radiation field of the LMXB. 
Note that in both models discussed above the requirement of a high space velocity suggests a relatively high proper motion, which may be detectable with future instrumentation.

\section{Conclusions}\label{sec:conclusions}
In a survey of faint persistent X-ray sources with \ch\ and EFOSC2 on the ESO 3.6m telescope, we find an accurate X-ray position of the 
neutron star low mass X-ray binary \thissource\, and identify its most likely optical counterpart: a faint source with blue colours with respect to field stars. Using the best known extinction and distance estimates, we find that a UCXB nature of  \thissource\  can be accomodated by the measured magnitudes. 
In the \ha\ data we serendipitously detect a large nebula with a distinct bow shock shape, likely associated with the LMXB. The nebula is likely formed through either the interaction of a energetic jet from the LMXB with the interstellar medium, or through intense UV/X-ray radiation. In both cases a high space velocity  is required,
and in both cases the nebula can provide unique insight into the energetics of this system.  
 
\section*{Acknowledgments}
Based on observations made with ESO Telescopes at the La Silla Observatory under programme ID 079.D-0239. 
We thank the anonymous referee, E.~Rol, J.~Miller-Jones, S.~Vaughan, T.~Maccarone, A.~Patruno and R. Starling for useful discussion. 
KW, AMR and NRT acknowledge STFC for support. DR acknowledges support from a 
%Netherlands Organization for Scientific Research (NWO) 
NWO Veni Fellowship.

\label{lastpage}

\end{document}